# Improved Mispronunciation detection system using a hybrid CTC-ATT based approach for L2 English speakers

Neha Baranwal and Sharatkumar Chilaka
IEEE Member and UpGrad
baranwal.neha24@gmail.com

**Abstract:** This report proposes state-of-the-art research in the field of Computer Assisted Language Learning (CALL). Mispronunciation detection is one of the core components of Computer Assisted Pronunciation Training (CAPT) systems which is a subset of CALL. Studies on automated pronunciation error detection began in the 1990s, but the development of full-fledged CAPTs has only accelerated in the last decade due to an increase in computing power and availability of mobile devices for recording speech required for pronunciation analysis. Detecting Pronunciation errors is a hard problem to solve as there is no formal definition of correct and incorrect pronunciation. As a result, typically prosodic and phoneme errors such as phoneme substitution, insertion, and deletion are detected. Also, it has been agreed upon that learning pronunciation should focus on speaker intelligibility rather than sounding like an L1 English speaker. Initially, methods were developed on posterior likelihood called Good of Pronunciation using Gaussian Mixture Model-Hidden Markov Model and Deep Neural Network-Hidden Markov Model approaches. These are complex systems to implement when compared with the recently proposed ASR based End-to-End mispronunciations detection systems. The purpose of this research is to create End-to-End (E2E) models using Connectionist Temporal Classification (CTC) and Attention-based sequence decoder. Recently, E2E models have shown considerable improvement in mispronunciation detection accuracy. This research will draw comparison amongst baseline models CNN-RNN-CTC, CNN-RNN-CTC with character sequence-based attention decoder, and CNN-RNN-CTC with phoneme-based decoder systems. This study will help us in deciding a better approach towards developing an efficient mispronunciation detection system.

**Introduction:** To establish the communication between human and machine[Baranwal et al, ] or human and human or machine and machine [singh et al, 2020, singh et al, 2019] gesture[Baranwal et al, 2017, Singh et al, 2018], speech, facial expression[singh et al, 2018] etc. are the important medium of communication where Speech(Baranwal et al, 2014, Baranwal et al, 2014) is the most natural form of human communication. There is a substantial rise in the market for learning foreign language with growing globalization, one among them is English pronunciation learning. Pronunciation teaching essentially involves a one-to-one interaction between pupil and teacher, which is unaffordable for many pupils. Hence, automated pronunciation teaching has become a popular research domain.

Study work on automatic identification of pronunciation errors and measurement of pronunciation began in the 1990s with a series of events from late 90s till early 2000s. The commercialization of CAPT at the beginning of 2000 proved problematic and, thus, development activities have slowed down. Interest started up again about thirteen years ago with increased computing capacity, smart devices and enhanced speech recognition. To understand the study better some of basic terminologies such as firstly Pronunciation and Types of Pronunciation errors need to be understood.

• Pronunciation error: Pronunciation is the style in which a phrase or a word is spoken. It is hard to calculate the word 'pronunciation error' because there is no definition available for correct and incorrect pronunciation. Instead, a complete spectrum is available from native-sounding speech remains to unintelligible speech.

• Types of Pronunciation errors

Errors in pronunciation can be classified into categories of phonemic and prosodic errors. The errors where phonemes may be replaced, omitted or added with another phoneme are phonemic errors. Errors in Non-native accent can be classified based on rhythm, intonation and stress on the prosodic side.

**Problem Statement:** Since the identification and teaching of pronunciation errors in their entirety is a difficult issue, previous study has mostly tackled only components of this area, such as the detection of pronunciation errors at phoneme level or the detection of prosodic errors.

Compared to the level of the syllable, word and phrase, a phoneme represents the smallest possible unit. The pronunciation quality judgement variability is higher for shorter units. The shorter the unit, the greater the uncertainty would be in pronunciation accuracy judgement.

Debate that began on the purpose of learning pronunciation concluded that training should be more focused on student's intelligibility rather than sounding like a native speaker. Although it is essential that advanced learners sound more like a L1 English speaker, obviously it is less essential than fundamental intelligibility. 3

There are two major industrial applications of pronunciation error detection: (1) as part of the measurement of pronunciation and (2) as part of the instruction of pronunciation. Each implementation presents its own challenges, especially on the pronunciation training side.

**Scope of the Study:** Currently, the focus of the research is limited to mispronunciation error identification in adult read speech only. The study can also be expanded to detecting mispronunciation errors in children's speech. This is especially useful to pupils belonging to rural areas of India, where shortage of teachers is major problem. A diverse country like India where each state has its own native language, coming up with a children speech corpus is time consuming and will go beyond the scope of this project.

Also, the experiments are being conducted on L2-Arctic, TIMIT datasets which are recorded in ideal conditions. When a CAPT system is used inside a mobile app, the recorded speech for evaluation would contain background noise as well. Currently this issue is not being addressed in this study.

**Analysis of previous research:** There was a great amount of research done for Mispronunciation detection techniques for developing efficient Computer Assisted Pronunciation Training systems. A complete CAPT system would consist of having methods to determine Mispronunciation detection and providing diagnosis feedback. (Witt, 2012) provides a detailed analysis on error detection in pronunciation training, wherein pronunciation errors are broadly classified into prosodic and phonemic errors. The phonemic errors are further sub divided into several other types. The characteristics used to detect errors in pronunciation are often derived from an HMM based speech recognizer's output. For phone level pronunciation scoring, (Witt and Young, 2000) present a modified version of the posterior likelihood technique called Goodness of Pronunciation. This technique is now commonly found in pronunciation assessment and identification of mispronunciation tasks. The methods mentioned in (Lo et al., 2010; Qian et al., 2010; Li et al., 2016) assess details of categorical pronunciation errors such as substitutions, insertions and deletions and provide diagnosis feedback about mispronunciations.

(Hu et al., 2015) builds three different methods based on Goodness of Pronunciation and two mispronunciation detection methods based on classifiers. The three GOP methods include one Gaussian Mixture Model-Hidden Markov based system and two Deep Neural Network-Hidden Markov Model based systems. DNN can effectively and efficiently model stress and tonal features which are highly unpredictable and need to be learnt for each word individually. Also, a revised GOP calculation is done to evaluate the non-native learner's pronunciation quality. The two classifier-based methods include Support Vector Machine based classifier and Neural Network based classifier. The shared hidden layer structure of the Neural Network not only increases the generality of certain phones that have fewer samples to create a strong classifier, but also helps in extracting more features to differentiate between right or wrong phones.

The transition probabilities heuristically ignore by most of the existing DNN-HMM based works. In the score calculation in the pronunciation assessment, the impact of these probabilities is not discussed. A new formulation for GOP under DNN-HMM dependent configuration is derived by (Sudhakara et al., 2019) using both senone posterior probabilities and probabilities of change. Furthermore, the paper also demonstrates the feasibility of the proposed formulation even when each senone is spread across several states by implementing it in Kaldi, a state-of-the-art open resource toolkit for automated speech recognition.

Compared to the methods, which rely primarily on the discovery of errors in categorical pronunciations such as phoneme substitutions, insertions and deletions, more emphasis is devoted to the detection of mispronunciations belonging to non-categorical errors or distortions in (Li et al., 2018).

(Lo et al., 2020; Yan et al., 2020; Zhang et al., 2020) propose the identification and diagnosis of mispronunciation with a specially customized E2E-based ASR model structure, where a hybrid CTC-Attention model is embodied in the involved E2E-based model. (Yan et al., 2020) also uses an anti-phone collection to generate additional speech training data with a label-shuffling scheme for a novel data-augmentation operation. The label of the phone at each point of its reference transcript is either kept unchanged or randomly replaced with an arbitrary anti-phone label for every utterance in the original speech training dataset.

**Research Gaps:** After going through the literature I found few research gaps which I discussed here.
• The system of CTC-ATT outperforms the GOP-based methods by a large margin, demonstrating the promise of using the model structure based on E2E for the task of detecting mispronunciation.
• An anti-phone approach is used by CTC-ATT to achieve better results. However, the diagnosis accuracy results of CTC-ATT, CTC, and ATT models are still less than expected, compared to using either CTC or ATT in isolation, using CTC-ATT stands out.
• This also means that the subtask for mispronunciation diagnosis is much more complex for the production of CAPT systems than the subtask for mispronunciation detection.

**Research Methodology:** Figure below shows the sequential steps involved in research experiments workflow. The process begins by preparing the data, then extracting fbank acoustic features. Next. Step would be train the CTC model and language model followed by decoding the results.

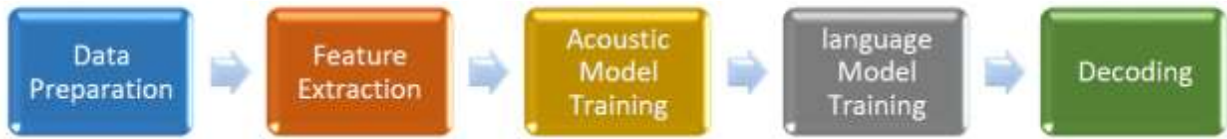

Figure1 Experiments workflow diagram

Kaldi Style Data Preparation:
Kaldi requires various transcript files to start extracting acoustic features from audio files.
1. Text: This file named "text" contains transcripts of all the utterances present in the speech corpus. It is text file with following format:utterance_id wrd1 wrd2 wrd3 wrd4 …
Data Preparation, Feature Extraction, Acoustic Model Training, language Model Training, Decoding Where, utterance_id is speaker id and file name of a particular audio sample appended together. Example of content inside the file: LXC_arctic_a0103 but there came no promise from the bow of the canoe
2. wav.scp: This file location of all the audio files present in the corpus. The content inside the text file has format as follows: utterance_id filelocation where, utterance_id is same as described above and filelocation is location of file on hard disk. Example of content inside the file:LXC_arctic_a0103/Users/sharat/L2-Arctic/LXC/wav/arctic_a0103.wav
Kaldi required the audio files to be in single channel .wav format. In case the audio files are in different format then it uses sox tool to convert them to wav format before extracting the features. The TIMIT corpus provides sphere headered file, hence the format of the content inside the file needs to be specified as shown below: FADG0_SA1 /Users/sharat/Desktop/FinalSubmission/kaldi/tools/sph2pipe_v2.5/sph2pipe -f wav /Users/sharat/timit/test/dr4/FADG0/SA1.WAV |
3. utt2spk: This is a text file that contains the mapping of each audio utterance to its corresponding speaker. It has the following format of content: utterance_id speaker_id where, utterance_id is the same as described above and speaker_id id the id/code name given to the speaker.
4. spk2utt: This is a text file contains the mapping of speaker to utterance. It has the following content format: speaker_id utterance_id where, utterance_id and speaker_id are the same as described in the utt2spk section. This file can be auto generated by reversing the information present inside utt2spk.
5. Segments: If corpus presents one audio file for each speaker which contains several utterance the we would need to create a segments file. This is a text file contains the start and end time for each utterance in the audio file. The content format is as follows: utterance_id file_name start_time end_time. Example of content format inside the file: LXC_arctic_a0103_1 arctic_a0103 13.0 22.0. This is an optional file and it is required only when there one audio with multiple utterances are present inside it.

The above mentioned files can be prepared via executing bash or python scripts. Kaldi also provides some example data preparation scripts for some of the commonly used speech corpora such as TIMIT, LibriSpeech, WSJ, Switchboard and many others.

Feature Extraction:
The next step in experiment pipeline is to extract acoustic features. After preparing the necessary transcript files, Kaldi feature extraction commands are executed. For our experiments purpose we will be using filter bank (fbank) features will be used. Filter bank feature work better than MFFC when deep neural networks are used since filter bank features are highly correlated features.
Below are the list of Kaldi library commands to compute fbank feature, compute cepstral mean, apply mean to features and copy features to fbank.scp file
• Example command to compute fbank feature:
compute-fbank-feats --config=conf/fbank.confscp,p:data/dev/wav_sph.scp ark:-
• Example command to compute cepstral mean and variance:
compute-cmvn-stats --binary=false scp:data/train/raw_fbank.scp data/global_fbank_cmvn.txt
• Example command to apply mean to features:
apply-cmvn --norm-vars=true data/global_fbank_cmvn.txt ark:- ark:-
• Example command to copy normalized features fbank features:
copy-feats --compress=false ark:-ark,scp:data/dev/fbank.ark,data/dev/fbank.scp

Acoustic Model Training:
Recently studies by (Amodei et al., 2015; Zhang et al., 2016) have shown that end-to-end models show good performance for ASR based tasks. Also in the literature review chapter we understood that they are simpler build and optimize as they do not have multiple modules to train separately when compared with conventional ASR systems. Studies done by (Leung et al., 2019; Feng et al., 2020; Yan et al., 2020) also showed very promising results for mispronunciation detection tasks and also outperformed the results obtained in (Li et al., 2017). The model proposed in this thesis is similar to (Feng et al., 2020) but instead of using sequence of characters for creating sentence embedding we utilize the sequence of phonemes to create sentence embedding. In this thesis project we will be using CTC loss function excluding time label information.

Connectionist Temporal Classification:
Connectionist temporal classification is a form of neural network output and associated scoring feature that is used to train recurrent neural networks such as LSTM networks to solve sequence problems with variable timing. CTC introduced in (Graves et al., 2006) refers to the outputs and scoring and is independent of the underlying neural network structure. It is a common option for tasks such as handwriting recognition and identifying phonemes in speech audio. A CTC network has a continuous performance that is fitted by training to model the possibility of a mark. The CTC scores will then be used for the back-propagation algorithm to change the neural network weights. Usually, CTC is a two-stage process: an acoustic model (encoder) followed by a translation model (decoder) at the level of letters (phonetic or word-piece) that produces sequential output by maintaining the left-to-right alignment order under a Markov assumption between the sequences of input and output (Graves et al., 2006).

Working of CTC:
CTC introduces a new symbol called blank symbol <b> to the existing set of characters/phonemes. At every time step the probability distribution of all characters/phonemes and the blank token are generated, the algorithm then selects the character/phoneme with highest probability. For example, consider the example where the target transcript to be generated is the word "cat". The CTC may generate a sequence of characters such as cccc<b>aaa<b>tttt<b> The basic rule for CTC is to collapse the repeated characters not separated by a blank. Hence the character sequence now gets collapsed to the word "cat". The sequences cc<b>aa<b>t<b>, cc<b>a<b>t<b>also collapse to form the word "cat".

Language Modelling:
A language model is built using the text available in the corpus. This is done to further improve the predicted output of CTC model. The CTC model may produce output transcript that is similar sounding to the input audio but it is not the correct expected transcription. For example consider below output transcript against the target transcription:
Target transcript:to illustrate the point a prominent middle east analyst in Washington recounts a call from one campaign
Output transcript: twoalstrait the point a prominent middle east analyst im Washington recouncacall from one campaign
Here the output transcripts sounded correct, but clearly lacked the correct spelling and grammar. These kinds of problems can be fixed by integrating a language model into CTC during training. During Kaldi setup additional software packages such as IRSTLM also get installed. IRSTLM is a language modelling tool for creating n-gram language models. Below are the example commands to create 2 gram language model using our training corpus.
build-lm.sh -i $srcdir/lm_train.text -n 2 -o lm_phone_bg.ilm.gz
compile-lm lm_phone_bg.ilm.gz -t=yes /dev/stdout> $srcdir/lm_phone_bg.arpa

Decoding:
An audio speech of length 10 seconds sampled at 100kHz generates 1000 input samples. The performance of an RNN falls drastically for longer sequences due to the problem of vanishing and exploding gradients. An attention model allows a neural network to pay attention to only a part of an input sequence while generating an output sequence. The output coming from audio encoder is passed attention decoder. The attention decoder makes use of beam search algorithm to find the best possible phoneme sequence as output.
Working details of attention mechanism: For each input text sentence, the Bidirectional LSTM encoder generates a sequence of concatenated forward and backward hidden states $(h_1, h_2, h_3, ... h_{N-1}, h_N)$ where, N is the number of words in input text sequence. During the creation of context vector (Bahdanau et al., 2014) focussed on the hidden states of all the words in the input sequence by taking a weighted sum of all the hidden states. The weighted sum of the annotations

(encoder output for each input time step) is used to generate the context vector $c_i$ for the output word $y_i$:

$$c_i = \sum_{j=1}^{N} \alpha_{ij} h_j \qquad (1)$$

where, $\alpha$ is a vector of (N, 1) dimension and of whose elements represent the weights assigned to words in the input sequence. For example if $\alpha$ is [0.3, 0.2, 0.2,0.3] and the input sequence is "How are you doing" the corresponding context vector here would be ci= 0.3∗h1 + 0.2∗h2 + 0.2 ∗h3 +0.3∗h4 where h1, h2, h3 and h4are hidden states (annotations) corresponding to the words "How", "are", "you" and "doing" respectively. Attention models are of two types Global attention and Local attention. In global attention model input from all hidden states must be taken into consideration, which results in increased computation. Now this happens because to obtain the final layer of the feedforward connection, all hidden states are added into a matrix and multiplied by a correct dimension weight matrix. To avoid this problem Local attention mechanism is used where only a few hidden states are considered.

Proposed Model:
Figure 2 shows the diagram of the proposed model structure. It helps understand the flow and organization of the model.

System Overview
The model takes two inputs, a phoneme sequence of transcript and filter bank (fbank) acoustic feature and outputs a corresponding audios phoneme sequence. Let x = (x1,x2,x3,...xt ,xt+1,...xT)represent the input audio, where $x_t$denotes the feature vector at frame t and T represents the total number of frames in the speech. Let s = (s1,s2,s3,...sn,sn+1,...sN) represent input phoneme sequence where sn is a phoneme at position n in the phoneme sequence. The model can be divided across three modules, sentence encoder, audio encoder and decoder with attention mechanism. The audio and sentence encoders extract high-level feature h.

$$\text{Audio encoder}(x) = h^Q \qquad (2)$$
$$\text{Sentence encoder}(s) = h^K, h^V \qquad (3)$$

$h^K, h^V$ and $h^Q$ are the fed as input to the attention decoder which then creates a context vector of fixed length.

$$\text{Attention}(h^Q, h^K, h^V) = cv \qquad (4)$$

Sentence Encoder
Each and every phoneme from the input phoneme sequence is embedded into a vector and fed to a Bidirectional Recurrent Neural Network (LSTM) which then outputs sequence($h_1^V, h_2^V, h_3^V$.................$h_N^V$), which is denoted as value. This value is fed into linear layer to obtain the output($h_1^K, h_2^K, h_3^K$.................$h_N^K$), which is denoted as key. The decoder module takes these key and value as input.

$$e_n = \text{Enbedded sentence }(s_n) \qquad (5)$$
$$h_N^V = f_1(e_n) \qquad (6)$$
$$h_N^K = f_2(h_N^V) \qquad (7)$$

Where f2(.)represents fully connected layer and f1(.) is function of Bidirectional RNN.

Audio Encoder:

The encoder is a sequence of CNN and RNN layers that takes an input feature of 243 dimension and generates the output sequence($h_1^Q, h_2^Q, h_3^Q, ... h_{t-1}^Q, h_{t'}^Q, ... h_{T-1}^Q, h_T^Q$) which is denoted as query. After the CNN down sampling operation the length of original T-frame becomes **T'**. Also, the original audio feature is 80 dimensional fbank and 1 dimensional energy with the stacking of left and right frames.

$$h_{t'}^Q = f_3(x_t) \qquad (8)$$

Where f3(.) is function of CNN-RNN layer.

Decoder with attention

The attention decoder creates a context vector cv which contains hK key information. The decoder makes use both future and past frames of a time sequence. Hence, a normalized weight learnt using $h_{t'}^Q$ and $h_N^K$ is given as:

$$\alpha_{t',n} = \frac{\exp\left(\text{score}(h_n^K, h_{t'}^Q)\right)}{\sum_{n=1}^N \exp\left(\text{score}(h_n^K, h_{t'}^Q)\right)} \qquad (9)$$

where, α is the attention weight that performs monotonic alignment between text input and the audio input. The context vector is computed as the weighted average of $h_n^V$ given as:

$$cv_{t'} = \sum_{t'=1}^{T'} \alpha_{t',n} h_n^V \qquad (10)$$

The output phoneme sequence $y_{t'}$ is generated using beam search algorithm.

$$y_{t'} = \text{softmax}(W[cv_{t'}; h_{t'}^Q] + b) \qquad (11)$$

where [·;·] represents concatenation of vectors.

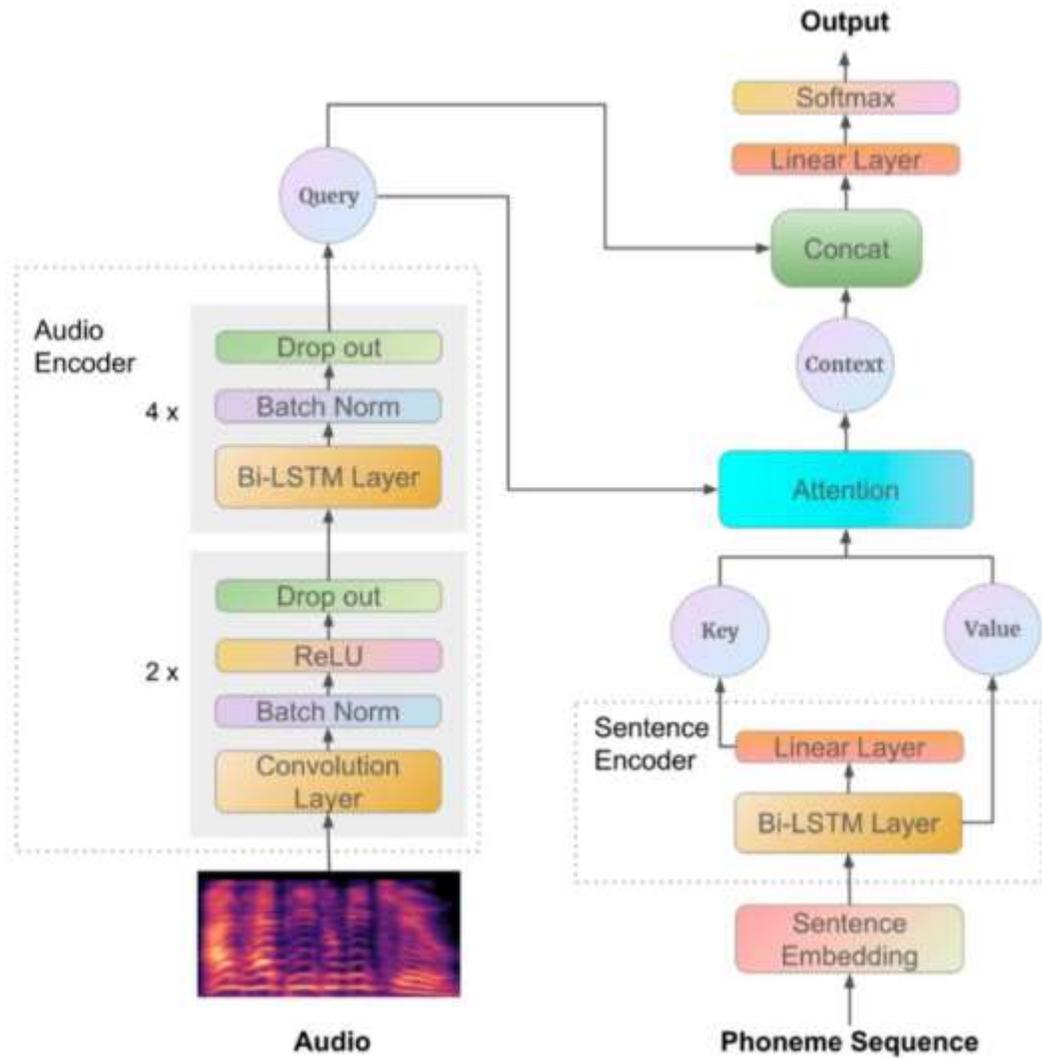

Figure 2 Proposed model diagram

**Evaluation Metrics**

The study proposes to compare the Precision, Recall and F1- score of baseline CNN-RNN-CTC model and CNN-RNN-CTC with Attention decoder models.

F-measure is defined as a harmonic mean of recall and precision which is defined as:

$$F1 - measure = 2 * \frac{Precision * Recall}{Precision + Recall} \qquad (12)$$

Using the confusion matrix for four test conditions illustrated below Precision and Recall metrics can be calculated.

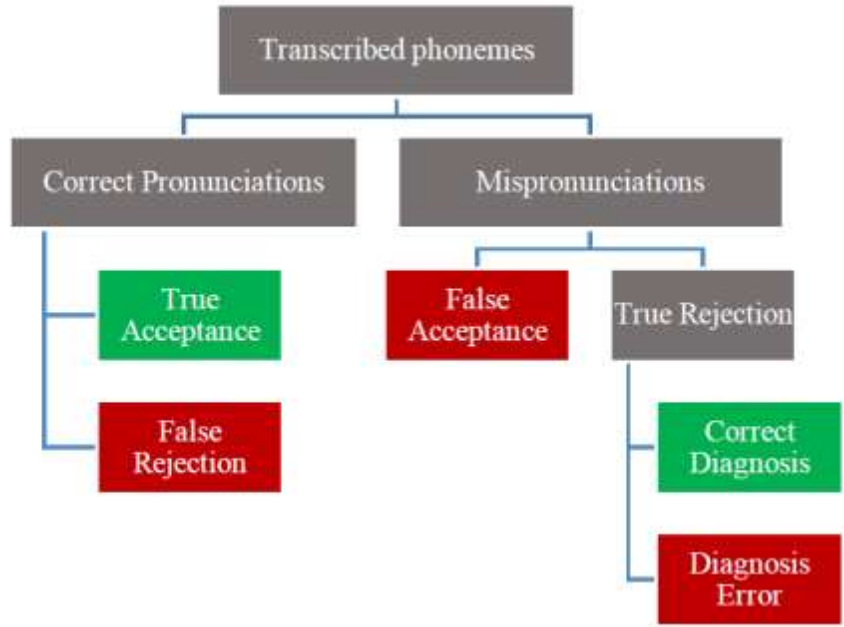

Figure 3 Hierarchical evaluation structure for mispronunciation detection

**Implementation:** The details of the two datasets TIMIT and L2-Arctic in terms of their purpose, content distribution, content type and content organization structure are discussed. We will also gain understanding of the L2-Arctic corpus statistics. Next data processing part will be discussed in terms of datasets are split for training and testing purpose, how it is prepared as per Kaldi style and how it will be augmented to solve the problem of data imbalance. The last section will provide details pertaining to model configuration.

1. Dataset description

For our experiments pertaining to our research project we use publicly available TIMIT(Garofolo et al., 1986) and L2 Arctic(Zhao et al., 2018) datasets. TIMIT is an old dataset designed for ASR tasks and L2-Arctic dataset is recorded of speaking CMU's Arctic prompts.

2. TIMIT dataset

The TIMIT speech corpus was created to offer data for the creation and assessment of ASR systems and extracting the acoustic features. TIMIT is the outcome of numerous locations working together under the support of the Defense Advanced Research Projects Agency - Information Science and Technology Office (DARPA-ISTO).The design of text corpora was a collaborative effort by the Texas Instruments (TI), Massachusetts Institute of Technology (MIT) and Stanford Research Institute (SRI).The National Institute of Standards and Technology (NIST) managed, verified, and prepared the speech for CD-ROM manufacturing after it was recorded at TI and transcribed at MIT.

3. Corpus speaker distribution

Table1 shows the number of male and female speakers selected for speech recording. TIMIT comprises a total of 6300 sentences, 10 of which are spoken by each of the 630 speakers from the United States' eight primary dialect regions.

Table1: TIMIT corpus speaker distribution

| Dialect region(dr) | Male | Female | Total |
|---|---|---|---|
| dr1 - New England | 31 | 18 | 49 |
| dr2 - Northern | 71 | 31 | 102 |
| dr3 - North Midland | 79 | 23 | 102 |
| dr4 - South Midland | 69 | 31 | 100 |
| dr5 - Southern | 62 | 36 | 98 |
| dr6 - New York City | 30 | 16 | 46 |
| dr7 - Western | 74 | 26 | 100 |
| dr8 - Army Brat | 22 | 11 | 33 |
| Total | 438 | 192 | 630 |

Table2 shows number of sentences selected for each speaker. The text material in the TIMIT prompts consists of 2 dialect "shibboleth" sentences (SA) designed at SRI, 450 phonetically compact sentences (SX) designed at MIT, and 1890 phonetically diverse sentences (SI) selected at TI. Below table summarizes the speech material in TIMIT.

Table2: TIMIT speech sentence distribution

| Sentence Type | No of Sentences | Sentences per speaker | Total |
|---|---|---|---|
| SA - Dialect | 2 | 2 | 1260 |
| SX - Compact | 450 | 5 | 3150 |
| SI - Diverse | 1890 | 3 | 1890 |
| Total | 2342 | 10 | 6300 |

**L-2 Arctic Dataset:** For experiments publicly available non-native English Speech corpus L2-Arctic is being used. It is intended in research for mispronunciation detection, accent and voice conversion. The corpus contains 26,867 utterances from 24 non-native speakers whose L1 languages are Arabic, Chinese, Hindi, Korean, Spanish and Vietnamese. The overall length of corpus is 27.1 hours and the average duration of speech per L2 user is 67.7 minutes. Over 238,702-word segments are included in the dataset, providing an average of about 9 words per utterance, and over 851,830 phone segments (Zhao et al., 2018).

Human annotators evaluated 3,599 utterances manually, annotating 1,092 phoneme addition errors, 14,098 phoneme substitution errors and 3,420 phone me deletion errors.

The corpus contains the following information for each speaker:
1. Speech recordings: more than one hour of recordings for phonetically balanced short sentences (~1132)
2. Word level transcriptions: for each sentence, orthographic transcription and forced-aligned word boundaries are provided.
3. Phoneme level transcriptions: for each sentence Montreal forced-aligned phonemic transcription are provided.

4. Manual annotations: a subset of 150 utterances are annotated with corrected word and phone boundaries, which includes 100 common utterances recorded for all speakers and 50 uncommon utterances that contain phonemes difficult to pronounce for each user as per their L1 language. These 150 utterances are tagged for phoneme substitution, deletion, and addition errors.

Every speaker's data is organized in its subdirectory under the root folder. Each speaker's directory is structured as follows:
• /wav: Containing audio files in WAV format, sampled at 44.1 kHz
• /transcript: Containing orthographic transcriptions, saved in TXT format
• /textgrid: Containing phoneme transcriptions generated from forced-alignment, saved in TextGrid format
• /annotation: Containing manual annotations, saved in TextGrid format

**Corpus speaker distribution:** Table3 shows the number sentences recorded annotated for each speaker as per their dialect and gender.

Table 3 L2-Arctic file summary and speaker information

| Speaker | Gender | Native Language | Number of Wav files | Number of annotations |
|---|---|---|---|---|
| ABA | M | Arabic | 1129 | 150 |
| SKA | F | Arabic | 974 | 150 |
| YBAA | M | Arabic | 1130 | 149 |
| ZHAA | F | Arabic | 1132 | 150 |
| BWC | M | Chinese | 1130 | 150 |
| LXC | F | Chinese | 1131 | 150 |
| NCC | F | Chinese | 1131 | 150 |
| TXHC | M | Chinese | 1132 | 150 |
| ASI | M | Hindi | 1131 | 150 |
| RRBI | M | Hindi | 1130 | 150 |
| SVBI | F | Hindi | 1132 | 150 |
| TNI | F | Hindi | 1131 | 150 |
| HJK | F | Korean | 1131 | 150 |
| HKK | M | Korean | 1131 | 150 |
| YDCK | F | Korean | 1131 | 150 |
| YKWK | M | Korean | 1131 | 150 |
| EBVS | M | Spanish | 1007 | 150 |
| ERMS | M | Spanish | 1132 | 150 |
| MBMPS | F | Spanish | 1132 | 150 |
| NJS | F | Spanish | 1131 | 150 |
| HQTV | M | Vietnamese | 1132 | 150 |
| PNV | F | Vietnamese | 1132 | 150 |
| THV | F | Vietnamese | 1132 | 150 |
| TLV | M | Vietnamese | 1132 | 150 |
| Total | | | 26867 | 3599 |

**Annotations**

The dataset uses ARPAbet phoneme series for the phonetic transcriptions as well as the error tags to make computer processing easier.

Each manually annotated TextGrid file will always have a "words" and "phones" tier while some of them may have an additional tier that contains comments from the annotators. The below are conditions for tagging a label to phone segment.
• For a correctly pronounced phoneme, the forced-alignment label remains unchanged.

- In case of a phone substitution error the forced-alignment label is replaced with following label template CPL,PPL,s, where CPL is the correct phoneme label (what should have been produced), PPL is the perceived phoneme label (what was actually produced) and s denotes for substitution error.
- When additional phone is present where it should not be, it is called as phone addition error and it is represented as sil,PPL,a where sil stands for silence and a stands for addition error.
- When a silent segment is found where there should be a phone segment, then it is called as phone deletion error and it is represented as CPL,sil,d where d denotes delete error.

**Exploratory data analysis**

An Exploratory Data Analysis (EDA) conducted on L2-Arctic dataset in (Zhao et al., 2018)for phoneme and pronunciation errors is shown in the following subsections. The results obtained by conducting EDA on L2-Arctic dataset helped in creating data augmentation techniques while training the model.

**Phoneme set distribution in L2-Arctic dataset**

Figure 4 shows the distribution of phonemes in the dataset. The phonemes 'AH', 'N', 'T', 'IH' and 'D' are top 5 phonemes found in the set.

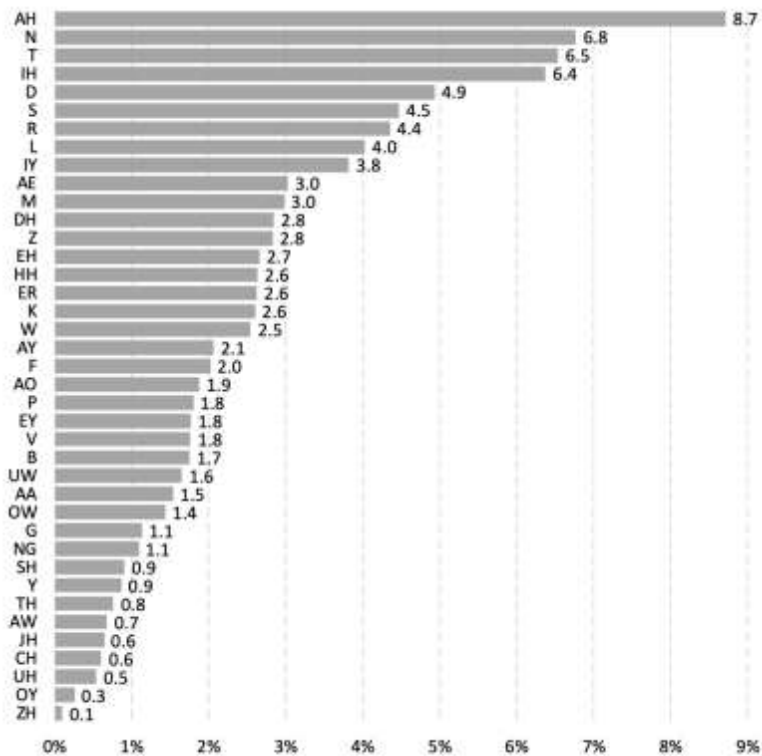

Figure 4 Distribution of phonemes in L2-Arctic dataset. (Zhao et al., 2018)

**Phoneme error distribution in L2-Arctic dataset**

Figure 5 show the top-20 most frequent phoneme substitution tags in the corpus. The most dominant substitution errors are 'Z->S', 'DH->D', 'IH->IY' and 'OW->AO'. Figure6 shows the phone deletion errors in the annotations. The most frequent phoneme deletions are 'D', 'T', and

'R'. Figure7 shows the phone addition errors in the annotations. The most frequent phoneme additions are 'AH', 'EH', 'R', 'AX', 'G' and 'IH'.

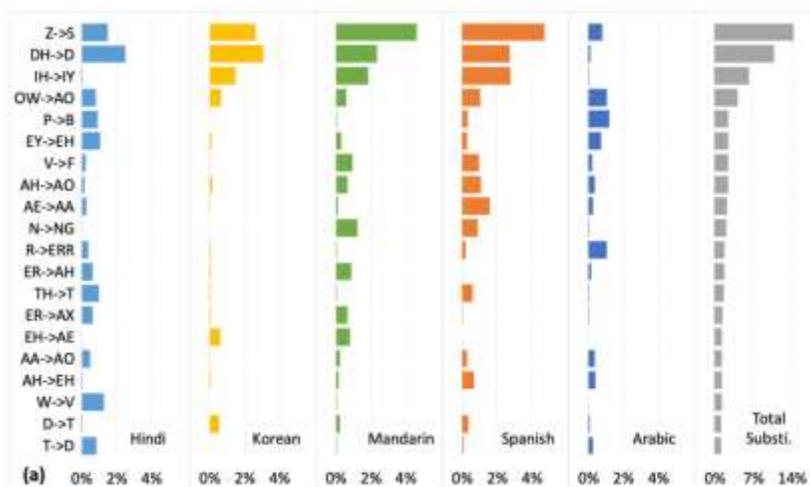

Figure 5 Top 20 most frequent dialect wise substitution errors in L2-Arctic. (Zhao et al., 2018)

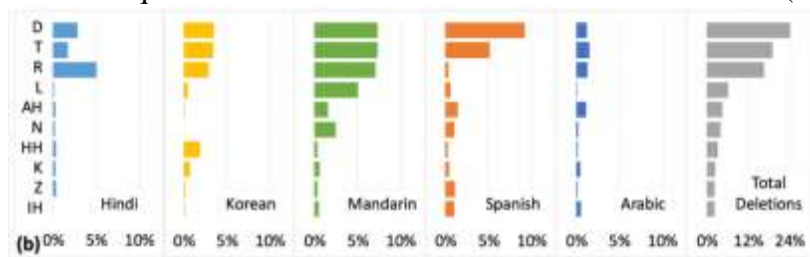

Figure 6 Dialect wise distribution of phoneme deletion errors. (Zhao et al., 2018)

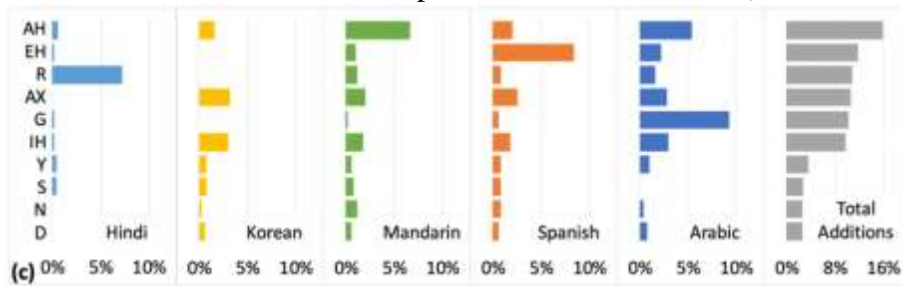

Figure 7 Dialect wise distribution of phoneme addition errors. (Zhao et al., 2018)

**Data processing**

The first step in the experiment begins by preparing data. As we will be using Kaldi toolkit, we need to follow Kaldi style data preparation as well.

**Data split**

The dataset required for training, validation and testing is split as shown in the Table 4.4. All the TIMIT utterances will be used for training. Out of the annotated utterances of L2-Arctic dataset, 12 speaker's utterances will be used for training, 6 speakers ("MBMPS", "YBAA", "NCC", "SVBI", "YDCK", "THV",) for validation and 6 speakers ("NJS", "ZHAA", "TXHC", "TNI", "YKWK", "TLV") for testing purpose. Total utterances used for training is 8100.

Table 4 Data split for training the model

|  | TIMIT | L2-Arctic | | |
| --- | --- | --- | --- | --- |
|  | Train | Train | Validation | Test |
| No of Speakers | 630 | 12 | 6 | 6 |
| No of Utterances | 6300 | 1800 | 897 | 900 |
| Total hours | 4.5 | 1.84 | 0.94 | 0.88 |

**Data Preparation**
TIMIT and L2-Arctic datasets have different directory structure and file types that hold the audio and transcripts data. Hence, both the datasets need to be prepared separately in their respective folders. For this purpose, separate scripts files are created and executed one after the other. The preparation phase essentially consists of creating Kaldi specific files that are
1. wav.scp: This file contains the location of all wav/audio files stored on disk for TIMIT train data this file will contain 6300 entries of wav/audio file locations. Similarly, there will be 3 different files for L2-Arctic dataset for training, validation and testing sets. Example: FADG0_SA1 /Users/sharat/timit/test/dr4/FADG0/SA1.WAV
2. wrd_text: This file contains the transcript for a given audio file, Example: FADG0_SA1 she had your dark suit in greasy wash water all year.
3. phn_text: This file contains the phonetic breakdown of trancripts for a given audio file Example: FADG0_SA1 silshiyhh ae sil d y er sil d aa r sil k s uwsil t ih n sil g r iy s iy w aoshsil w ao dh er ao l y ih er sil
After preparing these files for both datasets, the training files are merged. Next step is to generate filter bank features using Kaldi functions.

**Data Augmentation**
The training data does not contain enough samples of mispronounced tags, due to which there exists a class imbalance and the model will learn correctly pronounced phonemes better. To counter this issue data augmentation needs to be done to increase the mispronounced samples. This is performed based on the findings from the L2-Arctic dataset statistical analysis. Below are the techniques using which mispronounced tags are increased.

1. Random phoneme replacement based

A phoneme is randomly selected from input phoneme sequence, now this phoneme is replaced with a random phoneme from phoneme set. When a "blank" symbol is replaced by a phoneme it generates a DELETE type of error.

2. Consonant vowel based

Based on the phoneme error analysis on L2-Arctic it is found that vowels are more likely mispronounced with vowels and consonants are more likely mispronounced with consonants. Therefore, when a randomly selected phoneme is vowel then it is replaced with a vowel and when a randomly selected phoneme is consonant it is replaced with a consonant.

3. Confusing pair based

A dictionary of confused pairs of phonemes is created based on the findings from L2-Arctic dataset. The randomly selected phoneme is replaced with one of the phonemes from its list of confused phonemes.

**Model Implementation**

Two model architectures will be created CNN-RNN-CTC and CNN-RNN-CTC with attention-based decoder. The CNN-RNN-CTC architecture will be considered as the baseline model. As discussed in the proposed model section of chapter 3, there will be primarily 3 subunits in the model architecture Audio Encoder, Sentence Encoder and Decoder. The Audio encoder module consists of two CNN layers and four Bidirectional LSTM layers with batch normalization after each layer for better model convergence. In the sentence encoder phoneme embedded vector of size 512 is provided as input to. Bidirectional LSTM with hidden size = 384 and dropout = 0.2 Below is the model configuration details for Audio Encoder and Sentence Encoder.

Number of parameters 21246432
0 Sequential(
(0): LayerCNN(
(conv): Conv2d(1, 32, kernel_size=(3, 3), stride=(1, 2), padding=(1, 1))
(batch_norm): BatchNorm2d(32, eps=1e-05, momentum=0.1, affine=True, track_running_stats=True)
(activation): ReLU(inplace=True)
(dropout): Dropout(p=0.2, inplace=False)
)
(1): LayerCNN(
(conv): Conv2d(32, 32, kernel_size=(3, 3), stride=(2, 2), padding=(1, 1))
(batch_norm): BatchNorm2d(32, eps=1e-05, momentum=0.1, affine=True, track_running_stats=True)
(activation): ReLU(inplace=True)
(dropout): Dropout(p=0.2, inplace=False)
)
)
1 Sequential(
(0): BatchRNN(
(rnn): LSTM(1952, 384, bias=False, bidirectional=True)
(dropout): Dropout(p=0.2, inplace=False)
)
(1): BatchRNN(
(batch_norm): BatchNorm1d(768, eps=1e-05, momentum=0.1, affine=True, track_running_stats=True)
(rnn): LSTM(768, 384, bias=False, bidirectional=True)
(dropout): Dropout(p=0.2, inplace=False)
)
(2): BatchRNN(

41

(batch_norm): BatchNorm1d(768, eps=1e-05, momentum=0.1, affine=True, track_running_stats=True)
(rnn): LSTM(768, 384, bias=False, bidirectional=True)
(dropout): Dropout(p=0.2, inplace=False)

```
)
(3): BatchRNN(
(batch_norm):     BatchNorm1d(768,    eps=1e-05,    momentum=0.1,    affine=True, track_running_stats=True)
(rnn): LSTM(768, 384, bias=False, bidirectional=True)
(dropout): Dropout(p=0.2, inplace=False)
)
)
2 Embedding(42, 512)
3 LSTM(512, 384, batch_first=True, bidirectional=True)
4 Linear(in_features=768, out_features=768, bias=False)
5 Sequential(
(0): BatchNorm1d(1536, eps=1e-05, momentum=0.1, affine=True, track_running_stats=True)
(1): Linear(in_features=1536, out_features=43, bias=False)
)
6 LogSoftmax(dim=-1)
```

**Result and Analysis:**

This report discusses the results obtained by performing experiments on the proposed model CNN-RNN-CTC with phoneme sequence-based attention decoder. The chapter begins by exploring the phoneme error rate value obtained for baseline models and proposed models. Next, we explore the most frequently misrecognized vowels and consonants. After that we discuss the metric to evaluate Mispronunciation detection algorithm will be explored. The chapter ends by discussing the performance improvements achieved by the proposed model.

**Results of Phoneme Error Rate**

Phoneme error rate denoted (PER) is computed by aligning the annotated phoneme sequence and recognized phoneme sequence using the edit distance algorithm. The experimental results of phoneme recognition are shown in Table 5.1. It shows that the PER is 27.75% for the baseline CNN-RNN-CTC model. The PER improves for our proposed model that is CNN-RNN-CTC with attention decoder. For character sequence input with attention decoder PER comes as 20.66% and for phoneme sequence input with attention decoder PER is obtained as 16.06%, which concludes that the original baseline model does not perform well on this data. Further with data augmentation techniques viz Confusion pairs-based augmentation the PER goes to 15.48%. Table below shows the PER achieved for baseline model and proposed model with various data augmentation techniques. For representational convenience purpose the model will be addressed with below name terminologies.

Table 5 Phoneme error rate results of different models

| Models | Phoneme Error Rate % |
|---|---|
| Baseline model CNN-RNN-CTC (Leung et al., 2019) | 27.75% |
| Proposed model with character sequence | 20.66% |
| Proposed model with phoneme sequence | 16.06% |
| Proposed model with 10% random phoneme augmentation | 15.52% |
| Proposed model with 20% random phoneme augmentation | 15.96% |
| Proposed model with 10% confusion pair augmentation | 15.48% |
| Proposed model with 20% confusion pair augmentation | 16.20% |
| Proposed model with 10% vowel consonant augmentation | 15.58% |
| Proposed model with 20% vowel consonant augmentation | 16.33% |

**Most frequently misrecognized phonemes**

This section enables to gain insights about which vowels and consonants are frequently misrecognized. Table 6 and 7 shows the most frequently misrecognized vowels and consonants respectively. The proposed model with 10% vowel consonant augmentation outperforms the baseline model.

Table 6 Results of most frequently misrecognized vowels

| Models | aa | ah | ae | eh | ih | iy |
|---|---|---|---|---|---|---|
| Baseline model CNN-RNN-CTC (Leung et al., 2019) | 231 | 2035 | 630 | 544 | 1256 | 1097 |
| Proposed model with 10% vowel consonant augmentation | 310 | 2421 | 757 | 694 | 1567 | 1214 |

Table 7 Results of most frequently misrecognized consonants

| Models | d | dh | t | sh | s | z |
|---|---|---|---|---|---|---|
| Baseline model CNN-RNN-CTC (Leung et al., 2019) | 1067 | 125 | 1280 | 313 | 1457 | 228 |
| Proposed model with 10% vowel consonant augmentation | 1278 | 171 | 1433 | 322 | 1466 | 311 |

**Results on Performance of Mispronunciation detection**

The evaluation metrics are shown in Table 7 and Table 8.

Table 7 F-measure results of different models

| Models | Recall | Precision | F-measure |
|---|---|---|---|
| Baseline model CNN-RNN-CTC (Leung et al., 2019) | 74.78% | 36.76% | 49.29% |
| Proposed model with character sequence | 62.32% | 45.02% | 52.28% |
| Proposed model with phoneme sequence | 50.41% | 54.79% | 52.51% |
| Proposed model with 10% random phoneme augmentation | 54.93% | 55.72% | 55.32% |
| Proposed model with 20% random phoneme augmentation | 56.82% | 54.95% | 55.87% |
| Proposed model with 10% confusion pair augmentation | 54.86% | 56.07% | 55.46% |
| Proposed model with 20% confusion pair augmentation | 57.68% | 54.46% | 56.02% |
| Proposed model with 10% vowel consonant augmentation | 56.12% | 56.04% | 56.08% |
| Proposed model with 20% vowel consonant augmentation | 56.42% | 53.98% | 55.17% |

Table 8 Results of hierarchical evaluation structure

| Models | Correct Pronunciations | | Mispronunciations | | |
|---|---|---|---|---|---|
| | True Acceptance (TA) | False Rejection (FR) | False Acceptance (FA) | True Rejection | |
| | | | | Correct Diagnosis (CD) | Diagnosis Error (DE) |
| Baseline model CNN-RNN-CTC (Leung et al., 2019) | 78.53% | 21.47% | 25.22% | 64.57% | 35.43% |
| Proposed model with character sequence | 87.30% | 12.70% | 37.68% | 69.52% | 30.48% |
| Proposed model with phoneme sequence | 93.06% | 6.94% | 49.59% | 73.78% | 26.26% |
| Proposed model with 10% random phoneme augmentation | 92.72% | 7.28% | 45.07% | 75.31% | 24.69% |
| Proposed model with 20% random phoneme augmentation | 92.60% | 7.4% | 44.63% | 73.40% | 26.60% |
| Proposed model with 10% confusion pair augmentation | 92.83% | 7.17% | 45.14% | 75.45% | 24.55% |
| Proposed model with 20% confusion pair augmentation | 91.95% | 8.05% | 42.32% | 74.10% | 25.90% |
| Proposed model with 10% vowel consonant augmentation | 92.65% | 7.35% | 43.88% | 74.96% | 25.04% |
| Proposed model with 20% vowel consonant augmentation | 91.77% | 8.03% | 43.59% | 73.90% | 30.42% |

**Discussion**

The proposed model in this thesis CNN-RNN-CTC with phoneme sequence based attention decoder outperforms the model proposed in (Leung et al., 2019; Feng et al., 2020) in terms F-measure score. The model improves the F-measure obtained in (Leung et al., 2019) from 49.49% to 52.51%, which is also a slight improvement from F-measure obtained in (Feng et al., 2020). Using the data augmentation techniques, the F-measure further improves for proposed model with 10% vowel consonant augmentation. It achieves F-measure value of 56.08% which is an improvement of 6.79% when compared with baseline CNN-RNN-CTC model. In a reliable mispronunciation detection system, the True Acceptance (TA) value needs to be higher i.e., correct pronunciations getting detected correctly. Also, more mispronunciations should get detected therefore Correct Diagnosis (CD) value in True Rejection should be higher. Our model achieves a True Rejection value of 93.06% and Correct diagnosis value of 75.45% for model with 10% confusion pair augmentation technique. The phoneme error rate value of the proposed model also improves when compared with baseline model. The proposed model with 10% confusion pair augmentation achieves PER of 15.48%.

**Conclusion and Future Scope:**
**Conclusion and Discussion**

This thesis presents end-to-end model for mispronunciation detection using a hybrid CTC and attention-based decoder architecture. The proposed model does make use of forced alignment information, instead it only relies on the audio samples and the corresponding text and phoneme transcriptions. Due to this the model becomes easier to train model and tune the hyperparameters also. Moreover, the data augmentation techniques improved the shortcomings of the model as well. Experiments performed on TIMIT, and L2-Arctic datasets demonstrate that proposed model improves the score of F-measure and True Acceptance when compared with baseline models.

**Contribution to Knowledge**

The thesis work contributes to the field of Computer assisted language learning. The proposed model proves that end-to-end architectures can be used not only of Automatic speech recognition systems but for mispronunciation detection and diagnosis-based applications also. The performance measurements obtained for the proposed model can be used as a baseline for other modelling approaches as well. It is also found that using data augmentation techniques performance of the model does not degrade. Similar modelling techniques can be used for creating mispronunciation detection systems for other languages and other L2-English language users also.

**Limitations**

One of the biggest limitations was the unavailability of large corpus of speech data. License and Membership is required to download speech corpus from Language Data Consortium website belonging to University of Pennsylvania. Many of recipes in the Kaldi and ESPnet toolkit belong to speech corpuses available on this website. It is recommended to use large corpus of speech data with End-to-End models for generating effective performance. Due to less computation power available and time constraints large corpus was not used for acoustic model training. Also, the L2-Arctic dataset is an imbalanced dataset when the proportion of mispronounced and correctly pronounced phonemes are considered.

**Future Recommendations**

In future large corpus of speech data (1000 hours) such as LibriSpeech can be used for training the models. Also, possible usage of pre-trained models and unsupervised learning for mispronunciation detection can be explored. Considering manually annotating the speech data by linguists as a tedious and time-consuming process, additional data augmentation techniques need to be found. New modelling techniques can be investigated for detecting non-categorical and distortion-based mispronunciation errors. Developing a speech recognizer or mispronunciation detection system for children is extremely challenging as their voice and accent tends to vary greatly when compared with adults. The proposed model also can be used tested on various children speech corpus available on LDC website.

**Notes:** For more information about different applications of AI and deep learning techniques please go through these papers [1] [2] [3] [4] [5] [6] [7] [8] [9] [10] [11] [12] [13] [14] [15] [16] [17] [18] [19] [20] [21] [22] [23] [24] [25] [26] [27] [28] [29] [30] [31] [32] [33].